\documentclass[12pt, preprint]{aastex}
\usepackage{graphicx}
\slugcomment{accepted for publication in ApJL}

\begin{document}

\title{A Time Dependent  Leptonic Model for Microquasar Jets: Application 
to LSI +61 303}

\author{S. Gupta and M. B\"{o}ttcher}
\affil{Astrophysical Institute, Department of Physics and Astronomy,
Clippinger Hall 251B, Ohio University, Athens, OH 45701 -- 2979, USA}

\begin{abstract}
The Galactic high-mass X-ray binary and jet source (microquasar) LSI~+61~303 
has recently been detected at TeV $\gamma$-ray energies by the MAGIC telescope. 
We have applied a time-dependent leptonic jet model to the broadband spectral 
energy distribution and suggested (though not unambiguously detected)
orbital modulation of the very high energy $\gamma$-ray 
emission of this source. Our model takes into account time dependent electron 
injection and acceleration, and the adiabatic and radiative cooling of non-thermal 
electrons. It includes synchrotron, synchrotron self-Compton and external 
inverse Compton (with seed photons from the companion star), 
as well as $\gamma\gamma$ absorption of $\gamma$-rays by starlight photons. 
The model can successfully reproduce the available multiwavelength observational 
data. Our best fit to the SED indicates that a magnetic field of $B_0 \sim 5 \times 
10^3$~G at $\sim 10^3 \, R_g$ is required, and electrons need to be accelerated out 
to TeV energies ($\gamma_2 = 10^6$) with a nonthermal injection spectrum with a 
spectral index of $q = 1.7$, indicating the operation of acceleration mechanisms
beyond the standard first-order Fermi mechanism at relativistic or non-relativistic 
shocks. The orbital modulation of the VHE $\gamma$-ray emission can be explained solely
by the geometrical effect of changes in the relative orientation of the stellar
companion with respect to the compact object and jet as it impacts the position
and depth of the $\gamma\gamma$ absorption trough. 
Such a scenario predicts a trend of spectral hardening during VHE $\gamma$-ray low orbital phases.
\end{abstract}


\section{Introduction}
X-ray binaries with relativistic jets, or microquasars, have recently been 
established as a new class of $\gamma$-ray emitting sources. Two sources
detected by EGRET on board the {\it Compton Gamma-Ray Observatory} were
found to be spatially consistent with the locations of microquasars, namely
LS 5039 and LSI +61 303 \citep{hartman99}. Observations of $\gtrsim$ 250 GeV 
$\gamma$-rays from LS 5039 with the High Energy Stereoscopic System (HESS) have 
recently shown this high-mass source to be a TeV emitter \citep{aharonian05}, 
also strengthening its earlier tentative identification with the EGRET source 
3EG J1824-1514 \citep{paredes2000}. The second source, LSI +61 303, associated 
with the COS-B source 2CG 135+01 \citep{hermsen77, gt78} and the EGRET source 
3EG J0241+6103 \citep{kniffen97}, was recently observed with the Major Atmospheric 
Gamma-ray Imaging Chrenkov (MAGIC) telescope by \cite{albert06}. They detected 
variable $\gamma$-ray emission above 100 GeV over six orbital cycles, 
suggesting a periodic modulation on the time scale of the orbital period of $29.496$~d, 
though further observations are necessary to firmy establish the correlation 
between the VHE variability and the orbital period.
The strongest detections were not at the periastron, suggesting that the modulation
might not be related to a modulation of the accretion rate 
 \citep[as suggested
by, e.g.,][]{romero03,bosch04b},
but rather to geometrical effects. Geometrical 
effects causing an orbital modulation of the high-energy emission include 
the azimuthal-angle dependence of the $\gamma\gamma$ absorption of high-energy
emission by starlight photons \citep{boettcher05,dubus06} or of the Compton upscattering
of starlight photons by relativistic electrons in the microquasar jet \citep{dermer06}.
 \cite{dubus06} has performed a detailed analysis of the $\gamma\gamma$ 
absorption of VHE photons in the photon field of the stellar companion, taking
into account the finite size of the star and the eccentricity of the orbit for the 
case of VHE emission from the surface of a putative neutron star associated with
LSI +61 303. Note that he used a definition of the phase such that phase 0 
corresponds to the periastron passage, so that his results should be shifted 
by $\Delta\psi = 0.23$ (see below) when compared to the definition commonly 
used throughout the literature on this object, including this {\it Letter}.
He found that the $\gamma\gamma$ absorption depth is expected to show a 
pronounced maximum at phases just before the periastron passage, when
the compact object is located behind the stellar companion. The recent 
VHE detections, along with the observation 
of X-ray jet structures in several 
microquasars using $Chandra$ and $XMM-Newton$ \citep[e.g.,]{corbel02, tomsick02}, 
have re-ignited interest in jet models for the high energy emission from microquasars, 
analogous to the commonly adopted models for blazars \citep[for a recent review see, 
e.g.][]{boettcher02}. 

In the leptonic model of microquasars, Very High Energy (VHE) emission might most
likely originate near the base of the mildly relativistic jet. The soft photons 
from the companion star, the accretion disk as well as from jet synchrotron 
radiation can be Compton upscattered by the ultra-relativistic electrons in
the jet. Steady-state leptonic jet models of the $\gamma$-ray emission from microquasars
have been presented, e.g., by \cite{bosch04a,bosch04b}, and \citep{dermer06}. 
A time-dependent, broadband leptonic jet model was presented in \cite{gupta06}, 
(henceforth, Paper 1) where an analytical solution to the electron kinetic 
equation was presented, restricting the analysis to Compton scattering in the 
Thomson regime. This $Letter$ follows up on the analysis of Paper 1, now 
incorporating a full Klein-Nishina treatment of the Compton scattering as 
well as the angle dependence of the stellar radiation field. We then apply
the new model to broadband observations of LSI +61 303, iincluding the effect 
of the orbital modulation on the $\gamma\gamma$ absorption in the $>$100GeV 
range. In \S \ref{model}, we present a general outline of the model geometry 
and the radiative processes involved. \S \ref{application} shows the application 
to LSI +61 303, and we conclude with a brief summary in \S \ref{summary}.

\section{\label{model}Model Description}

The accretion flow onto the central compact object is ejecting a twin
pair of jets, assumed to be oriented perpendicular to the orbital plane,
which is inclined with respect to the line of sight by an angle $i$. 
Two intrinsically identical disturbances, containing non-thermal
plasma (blobs) originate from the central source at the same time, 
traveling in opposite directions along the jet at a constant speed 
$v_j = \beta_j \, c$. Over a limited range in distance $x^{}_0 \le x^{} 
\le x^{}_1$, relativistic electrons are accelerated and injected
in the emission region with an exponentially cut-off power-law distribution 
in electron energies ($E_e = \gamma \, m_e c^2$) with low- and high-energy
cutoffs $\gamma_1$ and $\gamma_2$, respectively, in the co-moving frame:

\begin{equation}
Q_{e}^{\rm inj}(\gamma;t) = Q_{0}^{inj}(t)\gamma^{-q} e^{-\gamma/\gamma_2},  
\;\;\;   \gamma_1\leq\gamma.
\end{equation} 
The normalization factor $Q_e^{\rm inj}$ is approximately related to the 
power injected into relativistic electrons, $L_{\rm inj}$ through

\begin{equation}
Q_0 \approx \cases{
      \frac{L_{\rm inj}(2-q)}{m_ec^2(\gamma_{max}^{2-q}-\gamma_{min}^{2-q})} & 
      \mbox{if $q \not= 2$};
\cr
      \frac{L_{inj}}{m_ec^2 ln(\gamma_{max}/\gamma_{min})} & \mbox{if $q=2$}. 
\cr}
\end{equation}

 The injection luminosity $L_{\rm inj}$ and the functional dependence
of Eq. (1) are held constant between $x_0$ and $x_1$.
The blob's (transverse) radius, $R_{\perp}$, scales with distance from the 
central engine as $R_{\perp} = R_{\perp}^0 \, (x / x_0)^{\alpha}$, i.e., 
$\alpha = 0$ corresponds to perfect collimation, and $\alpha = 1$ describes 
a conical jet. Following the arguments given in \cite{aa97}, we choose a 
magnetic-field dependence on distance from the central black hole as 
$B (x) = B_0 \, (R_{\perp}/R_{\perp}^0)^{-2} = B_0 \, (x / x_0)^{-2 \alpha}$. 

\begin{figure}[t]
\includegraphics[height=12cm]{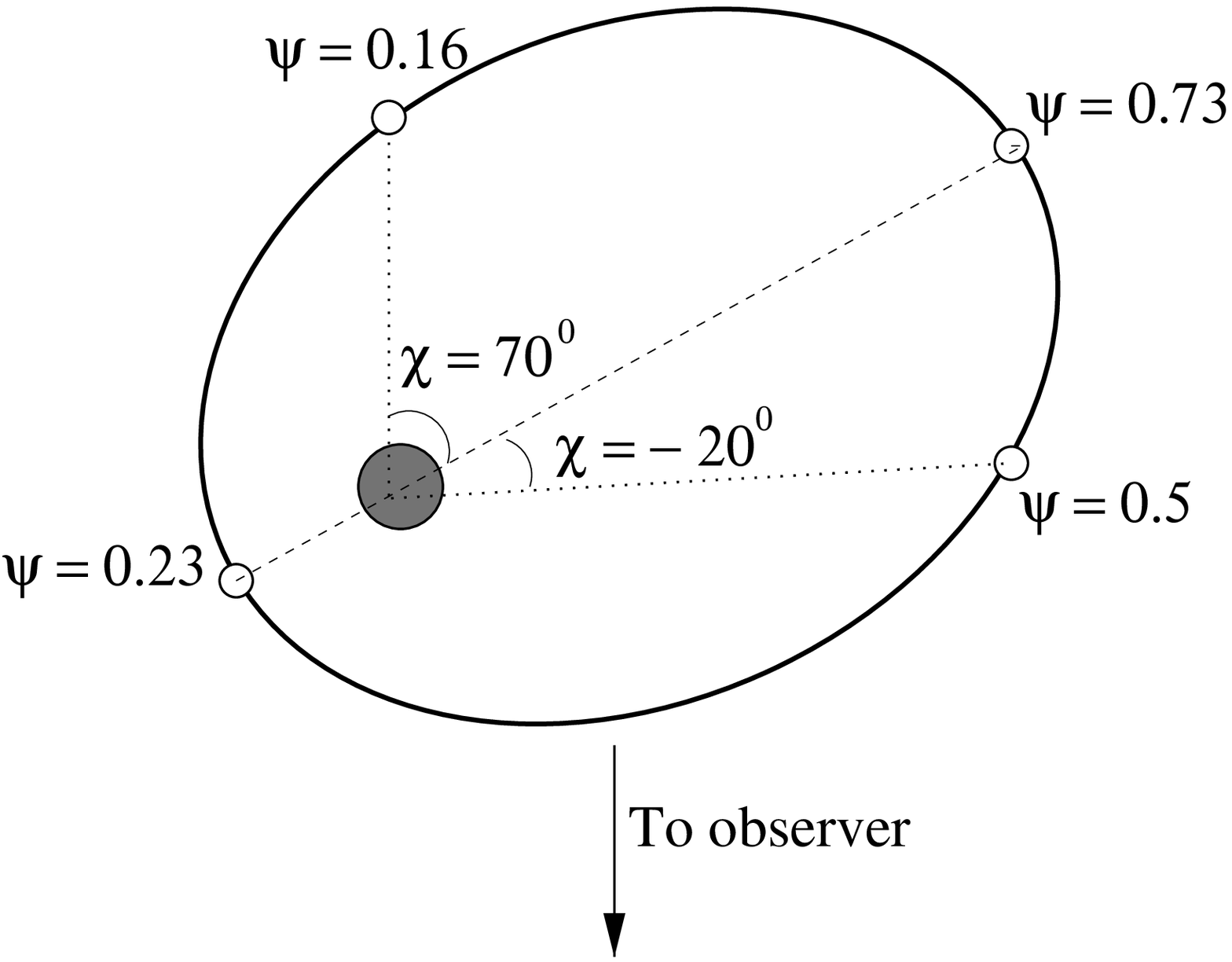}
\caption{Sketch of the geometry of the binary orbit of LSI +61 303, approximating
the massive (Be) star to be stationary. The labels on the compact-object orbit
indicate the orbital phase $\psi$ as well as the azimuthal angle with respect to
the apastron, $\chi$.}
\label{geometry}
\end{figure}

The geometry of the orbit of the LSI~+61~303 system is illustrated in 
Fig. \ref{geometry}. In order to incorporate the orbital modulation of 
the Compton scattering, we replaced the circular orbit of the binary 
system in Paper 1 by an elliptical one, so that the distance of the star 
to the compact object now varies as
\begin{equation}
r(\chi) = \frac{a(1-e^2)}  {1-e\cos(\chi)}
\label{r_chi}
\end{equation}
where $a$ is the semi-major axis,  $e$ is the eccentricity of the orbit, and 
$\chi$ is the phase angle \citep[which is {\it not} linearly related to the 
orbital phase $\psi$, in contrast to the incorrect expression in][]{romero05}. 
The $\gamma\gamma$ opacity as a function of $\gamma$-ray photon energy and 
orbital phase has been calculated following \cite{boettcher05}.

To treat the time-dependent electron dynamics and the radiation transfer in the
emitting blob, we solve the continuity equation for the relativistic electrons,
\begin{equation}
{\partial n_e(\gamma,t) \over \partial t}=-\frac {\partial}{\partial \gamma}\Big [
\Big ( \frac{d\gamma}{dt} \Big)_{loss} n_e(\gamma,t) \Big ] + Q_e^{\rm inj} (\gamma,t)
\end{equation}

where, $(d\gamma/dt)_{loss}$ is the 
 (radiative and adiabatic)
energy loss rate for the electrons.
The electron cooling rates are calculated using subroutines used in the jet
radiation transfer code of \cite{bms97} and \cite{boettcher2000}. The discretized
electron continuity equation can be written in the form of a tridiagonal matrix
\citep{chiaberge99}, and solved using the standard routine of \cite{press92}. 

Radiation mechanisms included in our simulations are synchrotron emission, 
Compton upscattering of synchrotron photons, namely synchrotron self-Compton
(SSC) emission, and Compton upscattering of external photons. 
 The dominant external soft photon source is the companion star
(for details, refer to Paper 1) which 
is approximated as a point
source, emitting a blackbody with dimensionless temperature
$\Theta_*=kT_*/m_{e}c^2$.

\section{\label{application} Application to LSI +61 303}

\subsection{\emph {Parameter Selection}}

The high mass XRB, LSI +61 303, at a distance of 2 kpc \citep{hutchings81, frail91}
has been observed from radio \citep{paredes96} to TeV $\gamma$-rays 
\citep{albert06}, providing a wealth of broadband data to test various emission 
models. However, most of the models until now dealt more with the radio, X-ray, or 
$\gamma$-ray variability of the source, than with its broadband properties and its
very-high-energy emission. Recently, however, \cite{chernyakova06} reproduced the 
broadband spectrum of the system, including the recent MAGIC detection, using a model 
of a rotation powered pulsar. The (slight) orbital modulation of the X-ray and 
$\gamma$-ray emission in the {\it XMM-Newton} and {\it INTEGRAL} energy ranges was 
explained based on a variable injection rate of high energy electrons. The VHE 
$\gamma$-ray emission was attributed to hadronic processes, initiated by high-energy 
protons in the pulsar wind, and its orbital modulation was not explained in that model.  
In contrast, in this {\it Letter}, we are specifically addressing the broadband
spectral characteristics {\it and} the orbital modulation of the VHE $\gamma$-ray
emission.

 \cite{casares05} have found a mass function of $f(M) \approx 0.011 \, M_{\odot}$.
Since the inclination angle $i$ between the line of sight and the normal to the
orbital plane (which, in our model is identical to the jet axis) is poorly 
constrained, and also the mass of the B0 V type stellar companion (with a 
dense equatorial wind) is not well known and in the range of $M_{\ast} \sim$ 
10 -- $15 \, M_{\odot}$, a rather wide range of masses for the compact object 
is consistent with the observational constraints, even including a $1.4 \, 
M_\odot$ neutron star \citep{hutchings81}. In our model, we follow the suggestion 
of \cite{casares05} of $i = 30^o$ and adopt a companion mass of $\sim 12M_\odot$, 
which would yield $M_X \sim 2.67 \, M_{\odot}$. The orbit has an eccentricity of 
$e=0.72\pm0.15$, the orbital period is $\sim$~26.496~d \citep{gregory02}, and the orbital semi 
major axis is $a = 5 \times 10^{12}$cm. MERLIN observations by \cite{massi04} 
imply a jet Lorentz factor of $\Gamma_{jet} = 1.25$. 

In our fitting procedure, we are varying the following parameters: the low
and high energy cut-off of the electron injection spectrum $\gamma_1$ and 
$\gamma_2$, respectively, the injection spectral index $q$, the initial distance 
of the injection zone from the compact object $x_0$, and the initial magnetic 
field $B_0$.  The straight power-law shape of the X-ray spectrum indicates that any 
disk blackbody component from an accretion disk around the compact object 
should have a luminosity of $L_D \lesssim 10^{34}$~ergs~s$^{-1}$ and be
at least $\sim 4$ orders of magnitude lower than the luminosity of the
companion star. For such low disk luminosities, external Compton scattering 
of accretion disk photons will be negligible as well. 
The evolution of
the electron distribution in the emitting region is followed over a period of 
$\sim$~5 days ($\ll P_{orb}$). Since most of the high-energy emission emanates 
from the system within the first few seconds of electron evolution, the orbital 
phase is taken to be a constant for each run of our model. Also, the centre of 
mass of the system lies close to the massive ($12 \, M_\odot$) stellar 
companion, so that the compact object essentially orbits the stellar companion,
which is assumed to be stationary. 

For an initial spectral fit, we concentrate on the orbital phase during which
MAGIC significantly detected the source. Specifically, we chose $\psi = 0.5$,
corresponding to a phase angle $\chi = -20^o$. This resulted in a best fit 
for which all physical parameters are summarized in Table ~\ref{par_table}. 

\begin{deluxetable}{lccc}
\tabletypesize{\scriptsize}
\tablecaption{Relevant parameter choices for our best fit to the broadband spectrum
of LSI~+61~303}
\tablewidth{0pt}
\tablehead{
\colhead{Parameter} & \colhead{Symbol} & \colhead{Value}
}
\startdata
Distance                                        & $d$                & $6.17 \times 10^{21}$~cm \\
Jet inclination angle                           & $i$                & $30^o$ \\
Bulk Lorentz factor                             & $\Gamma_j$         & $1.25$ \\
Semi Major Axis                                 & $a$                & $5 \times 10^{12}$~cm \\
Luminosity of companion star:                   & $L_{\ast}$         & $2\times 10^{38}$~ergs~s$^{-1}$ \\
Surface temperature of the companion star       & $T_{\ast}$         & $2.25 \times 10^4$~K \\
Initial blob radius                             & $R_0$              & $10^3 \, R_g$ \\
Jet collimation parameter                       & $\alpha$           & $0.3$ \\
Electron injection spectrum, low-energy cutoff  & $\gamma_{\rm min}$ & 10 \\
Electron injection spectrum, high-energy cutoff & $\gamma_{\rm max}$ & $10^6$ \\
Electron injection spectrum, spectral index     & $q$                & 1.7 \\
Beginning of electron injection zone            & $x_0$              & $10^3 \, R_g$ \\
End of electron injection zone                  & $x_1$              & $10^5 \, R_g$ \\
Magnetic field at $x_0$                         & $B_0$              & $5 \times 10^3$~G \\
Injection luminosity                            & $L_{\rm inj}$      & $10^{35}$~ergs~s$^{-1}$  \\
Orbital Period                                  & $P$                & $26.496$~days\\
Orbit eccentricity                              & $e$                & $0.72$ \\
\enddata
\label{par_table}
\end{deluxetable}

\subsection{\emph {Results}}

\begin{figure}[t]
\includegraphics[height=12cm]{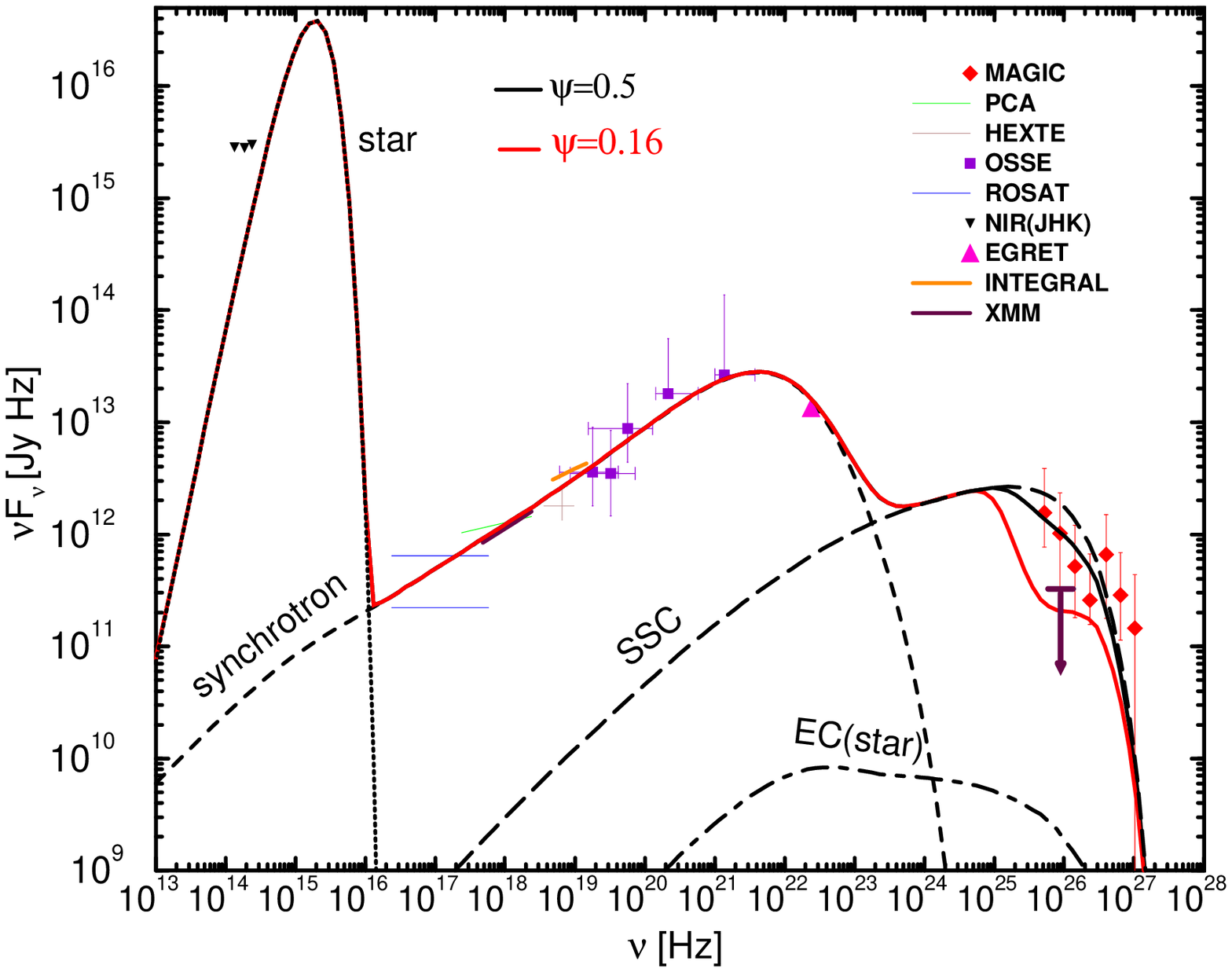}
\caption{Spectral energy distribution of LSI +61 303. The plotted observational data 
corresponds to $\psi \sim 0.5$ and are taken from: \cite{albert06} (MAGIC), 
\cite{harrison00} (RXTE PCA + HEXTE), \cite{strickman98} (OSSE, NIR), \cite{goldoni95} 
(ROSAT), \cite{leahy03} (EGRET), \cite{chernyakova06} (XMM, INTEGRAL). Also included
is the MAGIC $2 \, \sigma$ upper limit for phase $\psi \sim 0.2$ \citep{albert06}.
 The solid curves show our fit results with $\gamma\gamma$ absorption calculated for
$\psi \sim 0.5$ (black) and $\psi \sim 0.16$ (red). A trend of spectral hardening
at TeV energies during VHE low phases is predicted.
}
\label{fit}
\end{figure}

In Fig. \ref{fit}, we show the computed SEDs of LSI +61 303 using the parameters listed 
in Table ~\ref{par_table}. The black solid line corresponds to our best fit to the SED 
at the orbital phase of $\psi = 0.5$ $(\chi = -20^o)$. The figure also shows the individual 
contributions of the synchrotron, stellar, and SSC emission, as well as the inverse 
Compton (EC) scattering of starlight photons. 

The overall agreement of the simulated SED at $\psi = 0.5$ with the observed spectrum 
is very good, given that the multiwavelength data is not simultaneous. Since the 
radio emission most likely originates from regions of the jet further away from 
the compact object than the evolution modeled here, our model underpredicts the 
VLA data (not included in the figure). The NIR data \citep{strickman98} has been 
de-reddened using $N_H = 8.4 \times 10^{21} $~cm$^{-2}$ \citep{taylor96}. The infrared
emission might be dominated by emission from the outer parts of the circumstellar 
disk around the Be star which is not included in our model. 

In order to reproduce the keV -- MeV data, a magnetic field of $5 \times 10^3$~G is 
required at a distance of $\sim 1.7 \times 10^{9}$~cm from the compact object. In 
the X-ray band, most of the radiation is synchrotron emission from the jet 
\citep[also see, e.g.][]{paredes06}, with negligible contributions from SSC 
or EC. The SSC contribution dominates the spectrum at TeV energies. 
Note that the individual SSC component in  Fig. \ref{fit} does not include 
the $\gamma\gamma$ absorption feature, which is only applied to the total
emanating spectrum. 

After our fit to the $\psi = 0.5$ phase spectrum, we performed a model simulation with 
the same intrinsic jet parameters, but changed the orbital geometry to $\psi =
0.16$ $(\chi = 70^o)$, corresponding to the MAGIC low state. Since the EC (star)
component was rather insignificant, this will impact essentially only the
$\gamma\gamma$ absorption. The corresponding model fit is illustrated by the
solid red line in Fig. \ref{fit}. In this case, the $\gamma\gamma$ absorption
trough is deeper because (a) VHE $\gamma$-rays pass by the star more closely
than their nearest approach in the $\psi = 0.5$ geometry, and (b) starlight
photons intercept VHE photons at a more favorable angle for $\gamma\gamma$
absorption, leading to a lower energy threshold for this process. In this
phase, our model predicts an integrated flux over 400~GeV of $F (E > 400 {\rm GeV})
\approx 2.2 \times 10^{-12}$~ph~cm$^{-2}$~s$^{-1}$. This is in perfect agreement 
with the $2 \sigma$ upper limit of $F (E > 400 {\rm GeV}) < 3 \times 
10^{-12}$~ph~cm$^{-2}$~s$^{-1}$ given in \cite{albert06} and indicated by the
maroon arrow in Fig. \ref{fit}. Consequently, we find that the orbital 
modulation of the VHE emission in LS~+61~303 can be explained solely by 
the effect of the azimuthal asymmetry of $\gamma\gamma$ absorption
feature due to starlight photons.

\section{\label{summary} Summary and Conclusions}

We have applied a leptonic jet model to the broadband spectral energy distribution
and orbital modulation of the VHE $\gamma$-ray emission of the microquasar LSI~+61~303, 
taking into account time dependent electron injection and acceleration, and the 
adiabatic and radiative cooling of non-thermal electrons. Our model includes synchrotron, 
 synchrotron self-Compton and external inverse Compton emission (with seed photons 
predominantly from the companion star), 
as well as $\gamma\gamma$ absorption of $\gamma$-rays
by starlight photons. Compton scattering is treated using the full Klein Nishina cross
section and the full angular dependence of the stellar radiation field. 

The model can successfully reproduce the available multiwavelength observational data, 
including the most recent MAGIC detection in the TeV range \citep{albert06}. Our best
fit to the SED indicates that a magnetic field of $B_0 \sim 5 \times 10^3$~G at $\sim
10^3 \, R_g$ is required, and electrons need to be accelerated out to TeV energies
($\gamma_2 = 10^6$) with a nonthermal injection spectrum with a spectral index of 
$q = 1.7$. Such an injection spectrum can not be achieved by the first-order Fermi
mechanism at relativistic or non-relativistic shocks \citep[e.g.,][]{gallant99,achterberg01}
and therefore suggests that 2$^{nd}$ order Fermi acceleration \citep{virtanen05} and/or 
acceleration at shear boundary layers \citep{ostrowski00,stawarz02,rieger04} may play
a significant role in the acceleration of relativistic particles in microquasar jets.
In our model, the X-ray and $\lesssim 1$~GeV $\gamma$-ray emission is dominated by
synchrotron emission from the jet, while SSC emission dominates the VHE $\gamma$-ray
emission.

The suggested orbital modulation of the VHE $\gamma$-ray emission can be explained solely
by the geometrical effect of changes in the relative orientation of the stellar
companion with respect to the compact object and jet as it impacts the position
and depth of the $\gamma\gamma$ absorption trough \citep{boettcher05}. 
If this interpretation is correct, a general trend of spectral hardening
at TeV energies during VHE $\gamma$-ray low phases is predicted.
Although
additional effects of a varying accretion rate due to the highly eccentric orbit
of the binary system are likely to occur, they are not necessary to explain the
observed orbital modulation of the VHE emission of LSI~+61~303.

\acknowledgements{
 We thank the anonymous referee for useful comments and a quick
review, and C. D. Dermer, G. Dubus, and V. Bosch-Ramon for helpful comments and 
discussions.
This work was partially supported by NASA through INTEGRAL GO (Theory) 
grant award no. NNG~05GK59G.}

\end{document}